\documentclass[apjl]{emulateapj}
\usepackage[]{natbib} \usepackage{graphicx, amsmath} \bibpunct{(}{)}{,}{a}{}{,} \begin{document}
\def\bullet{\object{1E0657$-$56}}
\def\bbullet{\object{MACS~J0025.4$-$1222}}
\def\HST{{\it HST}}

\def\arcsecf{\!\!^{\prime\prime}}
\def\arcminf{\!\!^{\prime}}
\def\diff{\mathrm{d}}
\def\ngx{N_{\mathrm{x}}}
\def\ngy{N_{\mathrm{y}}}
\def\eck#1{\left\lbrack #1 \right\rbrack}
\def\eckk#1{\bigl[ #1 \bigr]}
\def\round#1{\left( #1 \right)}
\def\abs#1{\left\vert #1 \right\vert}
\def\wave#1{\left\lbrace #1 \right\rbrace}
\def\ave#1{\left\langle #1 \right\rangle}
\def\kms{{\rm \:km\:s}^{-1}}
\def\dds{D_{\mathrm{ds}}}
\def\dd{D_{\mathrm{d}}}
\def\ds{D_{\mathrm{s}}}
\def\cs{\mbox{cm}^2\mbox{g}^{-1}}
\def\magz{m_{\rm z}}
\def\V{\rm{V_{\rm 606}}}
\def\ii{\rm{i_{\rm 775W}}} 
\def\iii{\rm{I_{\rm 814W}}} 
\def\z{\rm{z_{\rm 850LP}}}
\def\J{\rm{J_{\rm 110W}}}
\def\H{\rm{H_{\rm 160W}}}
\def\chone{[3.6 \mu\rm{m}]}
\def\cunit{\mbox{erg}/\mbox{s}/\mbox{cm}^2/\mbox{\AA}}
\def\lunit{\mbox{erg}/\mbox{s}/\mbox{cm}^2}

\title{Spectroscopic Confirmation of a $z=6.740$ Galaxy behind the Bullet Cluster\altaffilmark{*}}
\altaffiltext{*}{Observations were carried out using the Very Large Telescope
at the ESO Paranal Observatory under Program ID 088.A-0542. Also based on
observations made with the NASA/ESA Hubble Space Telescope, obtained
at the Space Telescope Science Institute, which is operated by the
Association of Universities for Research in Astronomy, Inc., under
NASA contract NAS 5-26555 and NNX08AD79G. These observations are associated with
programs \# GO10200, GO10863, and GO11099.}
\shorttitle{}
\author{Maru\v{s}a Brada\v{c}\altaffilmark{1},
Eros Vanzella\altaffilmark{2},
Nicholas Hall\altaffilmark{1},
Tommaso Treu\altaffilmark{3,x},
Adriano\ Fontana\altaffilmark{4},
Anthony \ H. Gonzalez\altaffilmark{5},
Douglas \ Clowe\altaffilmark{6},
Dennis\ Zaritsky\altaffilmark{7},
Massimo\ Stiavelli\altaffilmark{8},
Benjamin Cl\'ement\altaffilmark{7}}
\shortauthors{Brada\v{c} et al.}
\altaffiltext{1}{Department of Physics, University of California, Davis, CA 95616, USA}
\altaffiltext{2}{INAF, Osservatorio Astronomico di Trieste, via G.B. Tiepolo 11, 34131 Trieste, Italy} 
\altaffiltext{3}{Department of Physics, University of California, Santa Barbara, CA 93106, USA}
\altaffiltext{4}{INAF, Osservatorio Astronomico di Roma, via Frascati 33, 00040 Monteporzio, Italy}
\altaffiltext{5}{Department of Astronomy, University of Florida, 211 Bryant Space Science Center, Gainesville, FL 32611, USA}
\altaffiltext{6}{Department of Physics \& Astronomy, Ohio University, Clippinger Labs 251B, Athens, OH 45701}
\altaffiltext{7}{Steward Observatory, University of Arizona, 933 N Cherry Ave., Tucson, AZ 85721, USA}
\altaffiltext{8}{Space Telescope Science Institute, 3700 San Martin Drive, Baltimore, MD 21218, USA}
\altaffiltext{x}{Sloan Fellow, Packard Fellow}
\email{marusa@physics.ucdavis.edu}


\begin{abstract}
  We present the first results of our spectroscopic follow-up of
  $6.5<z<10$ candidate galaxies behind clusters of galaxies. We report
  the spectroscopic confirmation of an intrinsically faint Lyman Break
  Galaxy (LBG) identified as a {$\z$}-band dropout behind the Bullet
  Cluster. We detect  an emission line at
  $\lambda=9412\mbox{\AA}$ at $>5$-$\sigma$ significance using a 16hrs
  long exposure with FORS2 VLT. Based on the absence of flux in bluer
  broad-band filters, the blue color of the source, and the absence of
  additional lines, we identify the line as Lyman-$\alpha$ at
  $z=6.740\pm0.003$. The integrated line flux is
  $f=(0.7\pm0.1\pm0.3){\times}10^{-17}\lunit$ (the uncertainties are
  due to random and flux calibration errors, respectively) making it
  the faintest Lyman-$\alpha$ flux detected at these redshifts.  Given
  the magnification of $\mu=3.0\pm0.2$ the intrinsic (corrected for
  lensing) flux is $f^{\rm int}=(0.23\pm0.03\pm0.10\pm0.02){\times}10^{-17}\lunit$ (additional uncertainty due to magnification),
  which is ${\sim}2-3$ times fainter than other
  such measurements in $z{\sim}7$ galaxies. The intrinsic {$\H$}-band magnitude of the
  object is $m^{\rm int}_{H_{\rm 160W}}=27.57\pm0.17$, corresponding
  to $0.5L^*$ for LBGs at these redshifts. The galaxy is one of
    the two sub-$L^*$ LBG galaxies spectroscopically confirmed at
    these high redshifts (the other is also a lensed $z=7.045$
    galaxy), making it a valuable probe for the neutral hydrogen
  fraction in the early Universe.  \end{abstract} 


\section{Introduction}
\label{sec:intro}

The epoch of reionization, which marks the end of the ``Dark Ages''
and the transformation of the universe from opaque to transparent, is
poorly understood.  It is thought that $z > 6$ faint proto-galaxies
were responsible for this transformation, but recent observations of
$z\gtrsim 7$ objects (e.g., \citealp{robertson10} for a review)
complicate that scenario. Finding robust samples of sources,
representative of the population contributing a significant
amount of energetic photons, is crucial.

Wide Field Camera 3 (WFC3) on HST enables a
preliminary identification of such galaxies. Substantial progress has
been made in detecting $z \gtrsim 7$ galaxies using the dropout
technique \citep{steidel96}, both in blank fields (HUDF, Candels, e.g.,
\citealp{bouwens12, oesch12,finkelstein12,mclure11}),
and behind galaxy clusters
(e.g., \citealp{kneib04,egami05,bradley12,richard11,zheng12,zitrin12}
and references therein).
One of the most obvious limitations of the dropout technique, however,
is that unambiguously confirming the object's redshift usually
requires spectroscopic follow-up. This is hard to do for typically
faint high-z sources, and it is thus an area where gravitational
lensing magnification helps greatly, as demonstrated in this paper.

In addition to the redshift confirmation, spectroscopy provides
information on properties of the interstellar and
intergalactic media (ISM and IGM).  In particular, Lyman-$\alpha$
emission from sources close to the reionization era is a valuable
diagnostic given that it is easily erased by neutral gas within and
around galaxies. Its observed strength in distant galaxies is a gauge
of the time when reionization was completed
\citep{robertson10}. Furthermore, we expect Lyman-$\alpha$
Emitters (LAEs) to be predominantly dust-free galaxies; hence their
numbers should increase with redshift until the state of the IGM
becomes neutral, at which point their numbers should decline. 

Significant progress has been made in detecting Lyman-$\alpha$
emitters (LAEs) in narrow band and spectroscopic surveys at $z\gtrsim
6$ (e.g., \citealp{kashikawa06,rhoads12,schenker12,clement12,
curtislake12,ono12, stark11, pentericci11}). Most studies see a
decline in the LAE population at $z>7$, but not all do
\citep{krug12, tilvi10}. The declining fraction of LAEs
within the LBG population \citep{stark10,kashikawa11,pentericci11} is
consistent with this decline being due to changes in the ISM/IGM,
specifically to an increased amount of neutral gas. However, current
studies only probe the bright end of the luminosity function of
LBGs. 

Furthermore, as noted by \citet{dijkstra11}, and \citet{dayal12}, measuring the
rest frame Equivalent Width (EW) distribution of LAEs as a function of
redshift {\it and} luminosity is a powerful tool to study
reionization. The EW distribution changes with redshift and source
luminosity. Simulations suggest that reionization is the key factor
driving this trend \citep{dayal12}, because unlike continuum photons,
the Lyman-$\alpha$ photons that escape the galactic environment
are attenuated by the \ion{H}{1} in the IGM.  With a
measurement of the EW distribution in LAEs we can therefore help
distinguish between effects of ISM dust and neutral IGM and study the
epoch of reionization (see also \citealp{treu12}). The main missing
observational ingredient is a measurement of the EW distribution for
both luminous and sub-$L^*$ galaxies at the redshifts of reionization
and this can only be achieved with spectroscopy.

Current spectroscopic observations unfortunately fall short of
matching the extremely deep near-IR {HST/WFC3} data for a significant
sample of $z\gtrsim 7$ dropout selected galaxies. Even with state of
the art facilities (e.g. the new spectrograph MOSFIRE on Keck,
\citealp{mclean10}) this will be a challenge. While samples of 
$\lesssim L^*$ galaxies at $z<6.5$ is steadily increasing (e.g.,\citealp{richard11, schenker12,labbe10}
for spectroscopic and imaging detections), to date, very few ${\lesssim}L^*$
galaxies at $z \gtrsim 6.5$ are spectroscopically
confirmed. The only examples are a lensed $z=7.045$
galaxy and a marginal detection at $z=6.905$ by \citet{schenker12}. At ${\sim}L^*$ a $z=6.944$ galaxy was detected by \citet{rhoads12}. At $z{\sim}8$
\citet{lehnert10} report a marginal detection of an emission line, but
independent observations do not detect it \citep[][in
prep.]{bunker12}. Other surveys
\citep{shibuya12, ono12,vanzella11,pentericci11,fontana10} target mostly brighter
sources. It is important to increase the sample at $z>6.5$ and compare it to
  $z<6.5$ because the timescale for changing the number of LAEs and their observed
EW distribution close to the reionization epoch is shorter than the
interval of cosmic time between $z{\simeq}6$ and $z{\simeq}7$
\citep{dayal12}.

A powerful way to detect emission lines from faint sources is to use
galaxy clusters as cosmic telescopes (e.g,\citealp{treu10} for a
recent review). Gravitational lensing magnifies solid angles
while preserving colors and surface brightness. Thus, sources appear
brighter than in the absence of lensing.
The advantages of cosmic telescopes are that we can probe
deeper (due to magnification), sources are practically always
enlarged, and identification is further eased if sources are multiply
imaged.  Typically, one can gain several magnitudes of magnification,
thus enabling the study of intrinsically lower-luminosity galaxies
that we would otherwise not be able to detect with even the largest telescopes. Indeed the highest
redshift sub-$L^*$ LBG currently spectroscopically confirmed is the
$z=7.045$ galaxy lensed by a cluster A1703 \citep{schenker12}. 
Observations using galaxy clusters as cosmic telescopes
are consistently delivering record holders in the search for the
highest redshift galaxies \citep{kneib04,bradley08,zheng12}. For this
reason we have started a large campaign of spectroscopic follow-up of
$z > 6.5$ candidates behind the best cosmic telescopes.  In this paper
we present the first spectroscopic confirmation from this campaign: a
$z=6.740\pm0.003$ galaxy behind the Bullet Cluster.

This paper is structured as follows. In Section~\ref{sec:data} we
describe the data acquisition and reduction, in
Section~\ref{sec:results} we present the spectrum of the galaxy and
we summarize our conclusions in
Section~\ref{sec:conclusions}. Throughout the paper we assume a
$\Lambda$CDM cosmology with $\Omega_{\rm m}=0.3$, $\Omega_{\Lambda}=0.7$, and Hubble constant $H_0=70{\rm\ kms^{-1}\:\mbox{Mpc}^{-1}}$.  Coordinates are given for the epoch
J2000.0, magnitudes are in the AB system.

\section{Targets, imaging and spectroscopic
  observations} \label{sec:data} 

\begin{figure*}[ht]
\begin{center}
\includegraphics[width=\textwidth]{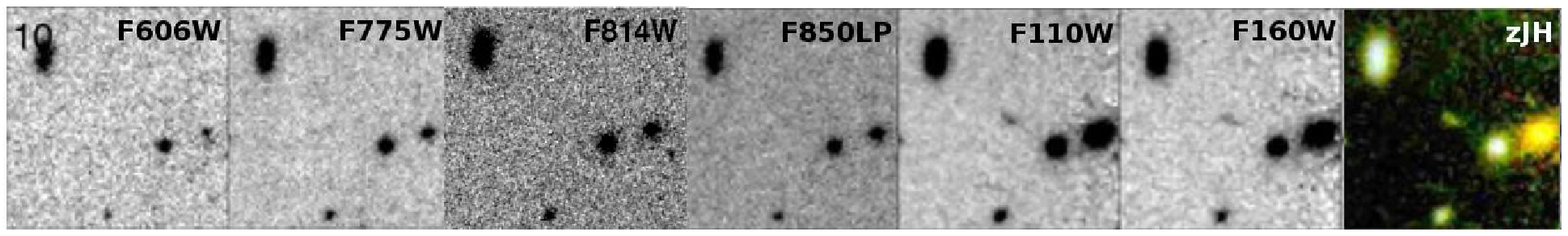} \\
\begin{tabular}{cc}
\includegraphics[width=0.5\textwidth]{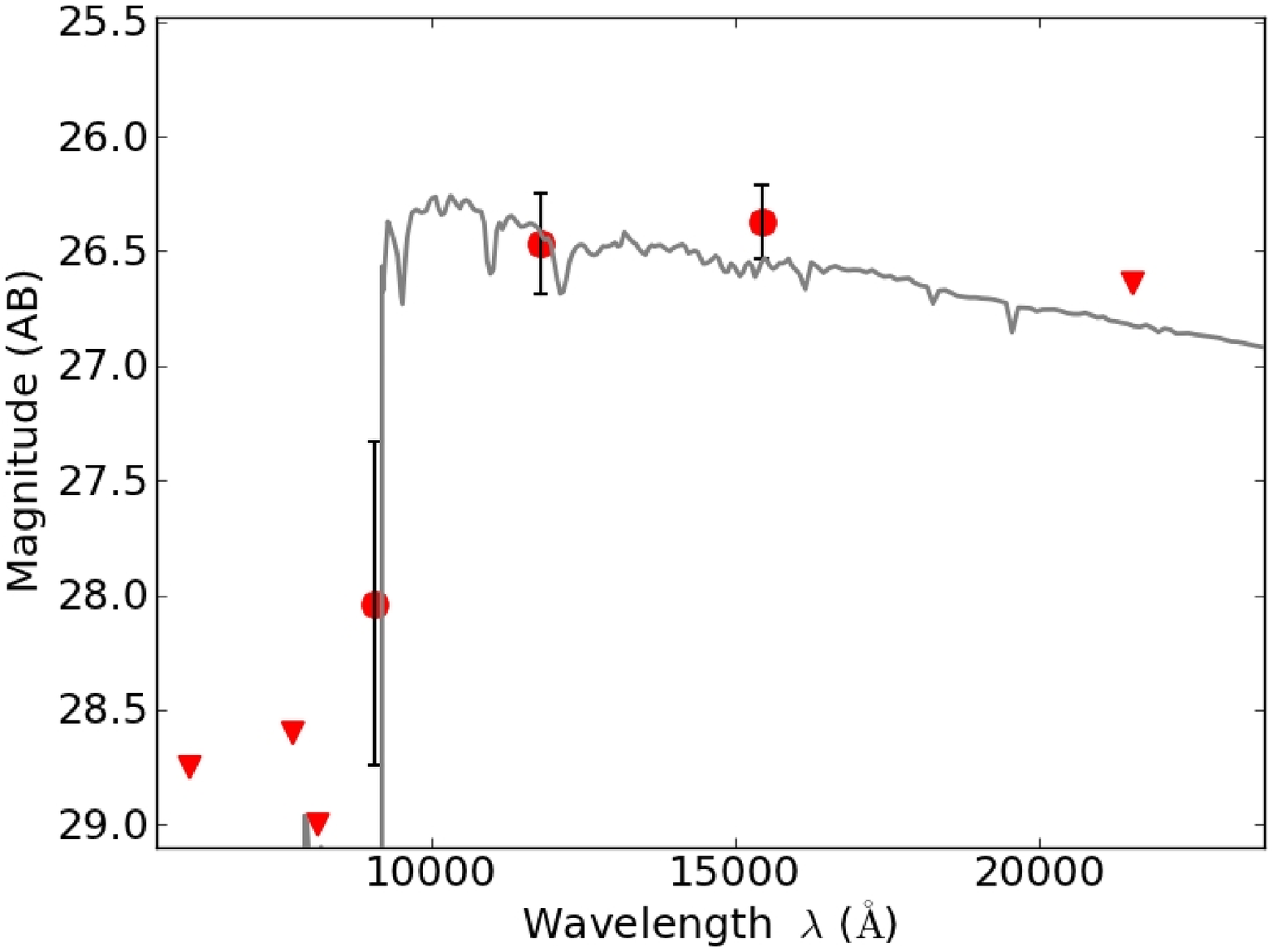} 
\includegraphics[width=0.5\textwidth]{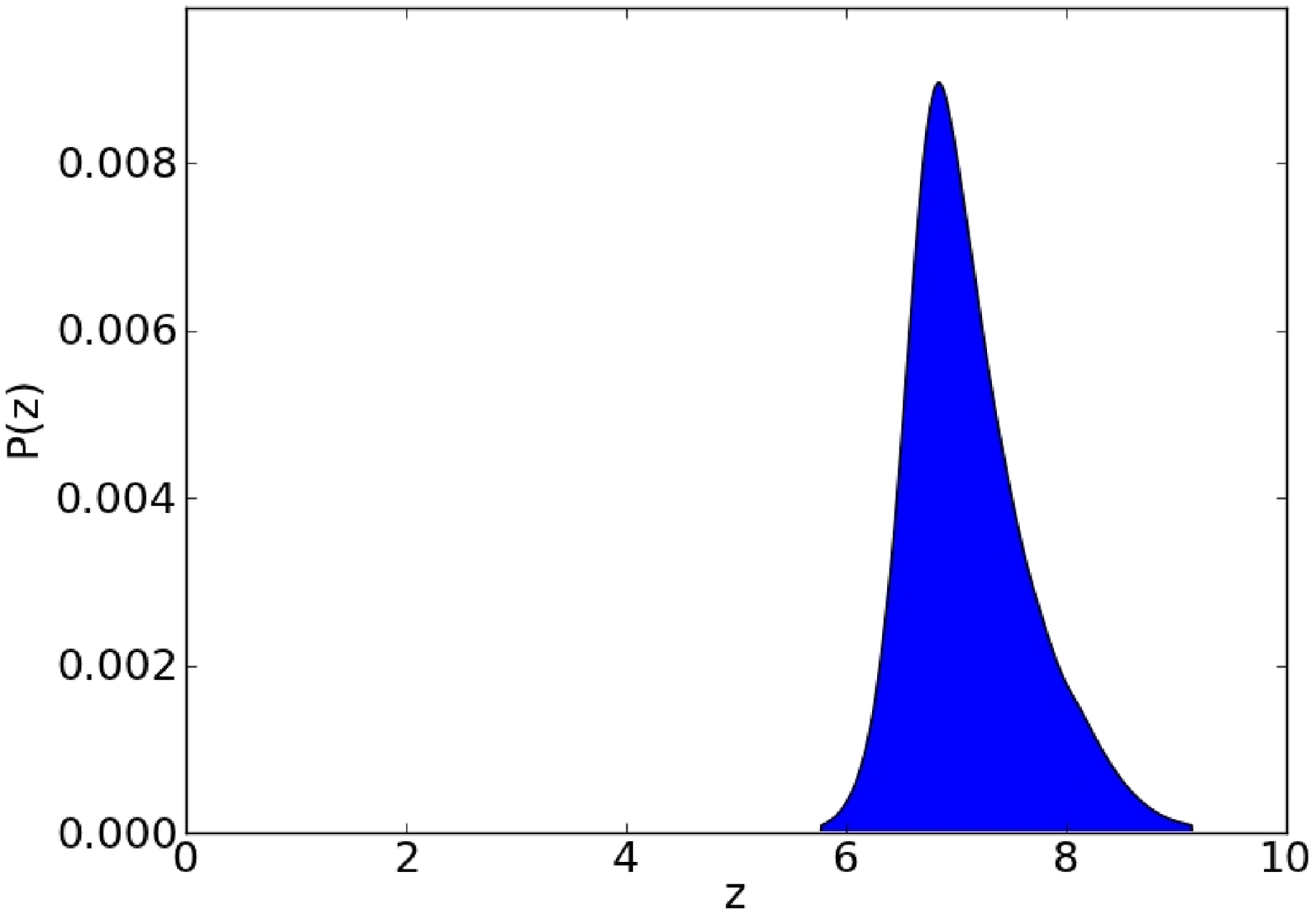} 
\end{tabular}
\end{center}
\caption{ ({\it top}) Cutouts of the dropout \#10 from
    \protect{\citet{hall12}} shown (from left to right) in $\V$,
    $\ii$, $\iii$, $\z$, $\J$, $\H$, {$\z\J\H$} color
    image (no bluer bands are available at present). The cutouts are $7\arcsec{\times}7\arcsec$, which
    corresponds to $20\mbox{kpc}$ at $z=6.740$ and magnification $\mu=3$. ({\it bottom-left}) BPZ SED fit to the photometry of the
    object. The photometric redshift is $z_{\rm
      phot}=6.8_{-0.8}^{+1.6}$ (95\% confidence). The circles give the
    observed AB magnitudes with uncertainties, while the triangles
    give the 1-$\sigma$ limiting magnitudes in cases of
    non-detection. ({\it bottom-right}) Posterior
  probability distribution of the photometric redshift. (bottom plots
  produced using BPZ tools, \citealp{coe10}).}
\label{fig:phot}
\end{figure*}
Our targets were selected from deep ACS/WFC3 HST
observations and are presented by \citet{hall12}. We targeted all 10
{$\z$}-band dropouts with the FORS2 spectrograph on the ESO Very Large
Telescope, Program ID 088.A-0542 (PI Brada\v{c}). Here we present
the first detection. The rest of the objects were not detected,
inferences obtained from the non-detections and the spectra of filler
slits will be presented in a subsequent paper.

The data were taken in service mode during 2011-2012
November-January. We used the 600Z holographic grating providing the
highest sensitivity in the spectral range of $8000-10000\mbox{\AA}$
with a resolution $R{\sim}1390$ and a sampling of $1.6\mbox{ {\AA}
  pixel}^{-1}$ for a $1\arcsec{\times}6 \arcsec$ slit. The spectra
presented here come from the coaddition of 42 exposures of 1400s of
integration each, with median seeing around $0.\arcsecf7$.  Series of
spectra were taken with two different masks, but all our main targets
were placed on both masks for a total integration time of
  16.3hr. Standard flat-fielding, bias subtraction, sky subtraction,
and wavelength calibration have been applied as in \citet{vanzella09,
  vanzella11}. To perform sky subtraction we fit a polynomial to
  the partial spectra extracted from the slit just above and below the
  target and apply the fit to the full spectrum. All the
sky-subtracted two-dimensional spectra (of a mask) were coadded in the
pixel domain. Finally, spectra were flux calibrated using observations
of spectrophotometric standards \citep{fontana10}. Slit losses are small, given the extremely compact
size of the targets and good seeing conditions, and have been
neglected in the subsequent discussion.

Most of the HST imaging data of the object are presented by
\citet{hall12} and summarized in Table~\ref{tab:drops}. In
  addition, we have reduced data from GO 11591 (PI Kneib) and do not
  detect the object in $\iii$. 
All 1-$\sigma$ limiting magnitudes are given in Table~\ref{tab:drops}.
We augment the HST data with imaging data from HAWK-I
\citep{clement12}. The object is undetected at 3-$\sigma$ in the 
Y, NB1060, J, and $\mbox{K}_{\rm s}$ bands, which is consistent with
our interpretation of the HST detections. For the purpose of
determining photometric redshift we only use the
HAWK-I non-detection in $\mbox{K}_{\rm s}$-band (the other bands
overlap with and are shallower than HST detections). The source is also
undetected in all four SPITZER IRAC bands ($\chone$,
$[4.5\mu\mbox{m}]$, $[5.8\mu\mbox{m}]$, and $[8\mu\mbox{m}]$) with
effective exposure times of $4\mbox{ks}$ in each filter. This is not
surprising given our detection limit for this object is $m_{H_{\rm
160W}}-m_{[3.6\mu\rm m]}=4$ at 3-$\sigma$. Unfortunately,
existing Spitzer images are too shallow to add to the quality of the
photometric redshift estimate, and we only use the data listed in
Table~\ref{tab:drops} in the Spectral Energy
Distribution (SED) fit (Fig.~\ref{fig:phot}).

For this we use the BPZ code \citep{benitez00} and assume
uniform priors on the spectral types and redshift. The observed SED is
best fit (reduced $\chi^2=0.3$) by a young starburst (5 Myr) galaxy model at $z_{\rm{phot}}=6.8_{-0.8}^{+1.6} $ (95\% confidence) which is in excellent
agreement with the spectroscopic data presented below
(Fig.~\ref{fig:spec}). The probability that the object is at low
  redshift given the set of templates used by BPZ is zero (Fig.~\ref{fig:phot}-right). Forcing the solution
  to $z<5$ the best fit SED is an elliptical galaxy template at
  $z_{\rm{phot}}=1.4_{-0.6}^{+0.7}$. To further discard the low-z
  solution would require either extremely deep F475W ($m_{\rm{F475W}}=30$), or deeper Spitzer data; neither of which exists for
  this object. However, the reduced $\chi^2$ of the fit for $z<5$
  increases by a factor of 10 compared to the high redshift solution;
  in addition the object is unlikely an elliptical galaxy given the
  presence of an emission line (see below). We
  therefore conclude that the object is most likely at $z>5$.

We also performed an independent fit to the photometric data using the
HyperZ code \citep{bolzonella00}, which has the advantage of treating
internal extinction as a free parameter. The results are very
  similar and the probability that the object
  is at high redshift is $>99\%$. Our conclusions are therefore robust even to
  cases of heavily reddened, dusty star forming galaxies like the one
 in \citep{gonzalez10}.

Finally, as noted in \citet{hall12} it is difficult to estimate
the size of the object given the uncertainty in measuring low surface
brightness objects. The object is resolved and we estimate the FWHM of
${\sim}0.\arcsecf26$ in $\J$ and ${\sim}0.\arcsecf21$ in $\H$. In
physical units and correcting for lensing this translates to ${\sim}0.8\mbox{kpc}$ which is
consistent with the compact sizes reported at these high redshifts
(e.g., \citealp{oesch10b}).

\section{Results}\label{sec:results}
\begin{deluxetable}{lc}
 \tablecolumns{2}
\tablewidth{0pc}
 \tablecaption{Imaging and spectroscopic properties of z$_{850}$-band dropout  \# 10 from \protect{\citet{hall12}}}
\startdata
\hline R.A. & 104.63015 \\ 
 Dec. & -55.970482\\ \hline
 $m_{H_{\rm 160W}}$ & 26.37$\pm$0.16\\ 
  $m_{J_{\rm 110W}}$& $26.5\pm0.3$\\
 $\left(J_{\rm 110W}-H_{160W}\right)$ &  0.10$\pm$0.15\\ 
 $\left(z_{\rm 850LP}-J_{\rm 110W}\right)$ &  1.57$\pm$0.68 \\ 
$m_{\V}$\tablenotemark{(a)}& $> 28.75$ ($t_{\rm exp} = 2336\mbox{s}$)\\
$m_{\ii}$ & $> 28.60$ ($t_{\rm exp} = 10150\mbox{s}$)\\
$m_{\iii}$ & $> 29.00$ ($t_{\rm exp} = 4480\mbox{s}$)\\
$m_{\rm K_s}$ & $> 26.65$ ($t_{\rm exp} = 3.75\mbox{hr}$)\\
\hline
 $\mu$ & 3.0$\pm$0.2\\ 
 $m^{\rm int}_{H_{\rm 160W}}$ & 27.57$^{+0.17}_{-0.17}$\\ \hline
$\lambda $ & $9412\mbox{\AA}$\\ 
$z$ & $6.740\pm0.003$\\ \hline
$f$\tablenotemark{(b)}& $(0.7\pm0.1\pm0.3){\times}10^{-17}\lunit$\\ 
$f_{\lambda,\rm{c}}$ &  $3.3_{-0.8}^{+1.0}{\times}10^{-20}\cunit$\\ 
$f^{\rm{int}}$& $(0.23\pm0.03\pm0.10\pm0.02){\times}10^{-17}\lunit$\\ 
$f^{\rm{int}}_{\lambda,\rm{c}}$ & $1.1_{-0.3}^{+0.4}{\times}10^{-20}\cunit$ \\ 
$W^{\rm{rest}}(\mbox{Ly}\alpha)$& $ 30^{+12}_{-21}\mbox{\AA}$
\enddata
\tablenotetext{(a)}{All upper limits are 1-$\sigma$ limiting magnitudes
  calculated in $0.63\arcsec{\times}0.63\arcsec$ square
  apertures. $t_{\rm{exp}}$ is the exposure time in corresponding band.}
\tablenotetext{(b)}{ $f$ is the integrated line flux, $f_{\lambda,c}$ is
  the continuum flux estimated from the $J_{\rm{110W}}$ magnitude,
  while {\it int} denotes corresponding intrinsic
  (unlensed) values. 
}
\label{tab:drops}
\end{deluxetable}

\begin{figure*}[ht]
\begin{center}
\includegraphics[width=\textwidth]{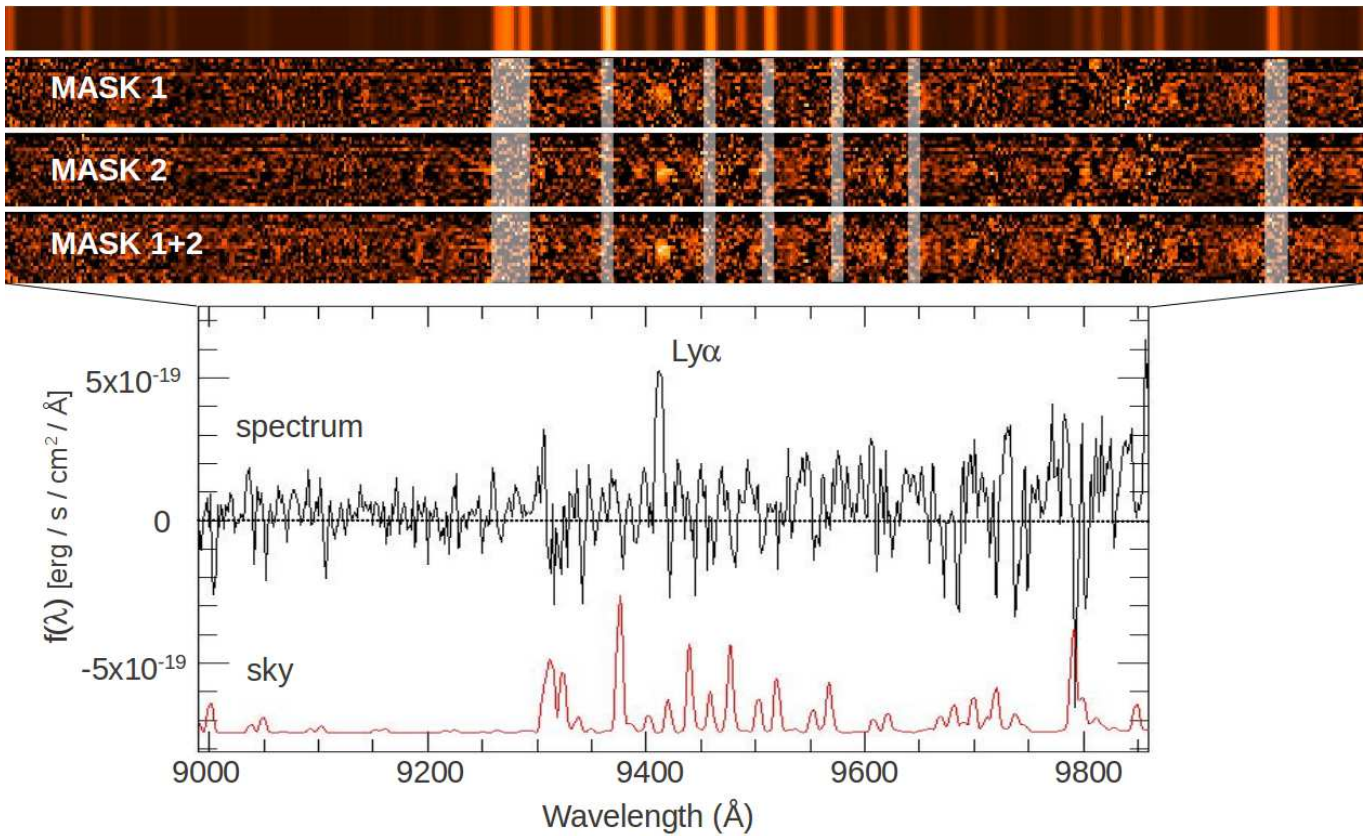} 
\end{center}
\caption{{\it (top)} 2-D spectrum of the dropout galaxy (sky emission
  above, spectra from individual masks below). {(bottom)} 1-D spectrum
  of the object. The sky spectrum has been rescaled by a factor of 300
  and offset for plot purposes, and the regions where skylines are
  more intense have been marked with transparent vertical bars. Strong
  residuals in the sky subtraction are evident and correspond to the
  more intense sky lines.  Note that the detected line is broader than
  residuals of sky subtraction, confirming the reality of the
  feature.}
\label{fig:spec}
\end{figure*}

The object we present here is the only one for which an emission line
was detected, out of the 10 $z$-band dropout candidates listed in
\citet{hall12}. We detect an emission line at $9412\mbox{\AA}$ with
$>5\mbox{-}\sigma$ significance (Figure~\ref{fig:spec}). The line
is detected in two different masks and is broader than cosmic rays or
residuals due to sky subtraction, hence we are confident that the line
is not an artifact. The integrated flux of the line is
$f=(0.7\pm0.1\pm0.3){\times}10^{-17} \lunit$, where the first error
corresponds to statistical uncertainty in the detection. The second
(larger) is due to systematic uncertainty in absolute flux calibration
and due to flux losses in the proximity of skylines and was estimated
based on previous multiple observations of various standard stars. No
other emission lines are detected in the spectrum
($7700\mbox{\AA}-10000\mbox{\AA}$), which one would expect for some of
the possible low-z solutions as discussed below.  Based on this and on
the SED fit we exclude other alternative explanations and conclude
that the line is most likely Lyman-$\alpha$ at $z=6.740\pm0.003$. This
agrees extremely well with the peak redshift probability distribution
described in Sect.~\ref{sec:data}.

Because of the relatively low signal-to-noise ratio of the spectrum we
cannot determine whether the line is asymmetric (which is
expected for high redshift LAEs where absorption happens mostly in the
blue wing of the line) and thus further test our identification as
Lyman-$\alpha$. However, we use indirect arguments to test the
alternative hypothesis that this galaxy is at a low redshift and the
line is not  Lyman-{$\alpha$} but: (1)
[\ion{O}{2}] ($3727\mbox{\AA}$) at $z=1.525$; (2) [\ion{O}{3}]
($4959\mbox{\AA},5007\mbox{\AA}$) at $z=0.898,0.880$; or (3)
H-$\alpha$ ($6563\mbox{\AA}$) at $z=0.434$, which are the 
prominent lines in emission line galaxies (e.g., \citealp{straughn09}).

\begin{figure*}[ht]
\begin{center}
\begin{minipage}{\textwidth}
\begin{minipage}{0.5\textwidth}
\includegraphics[width=\textwidth]{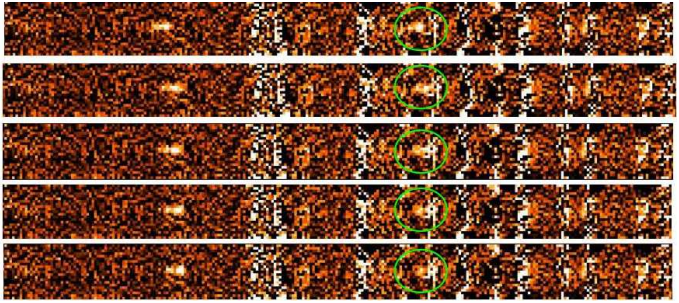} 
\end{minipage}
\begin{minipage}{0.5\textwidth}
\includegraphics[width=\textwidth]{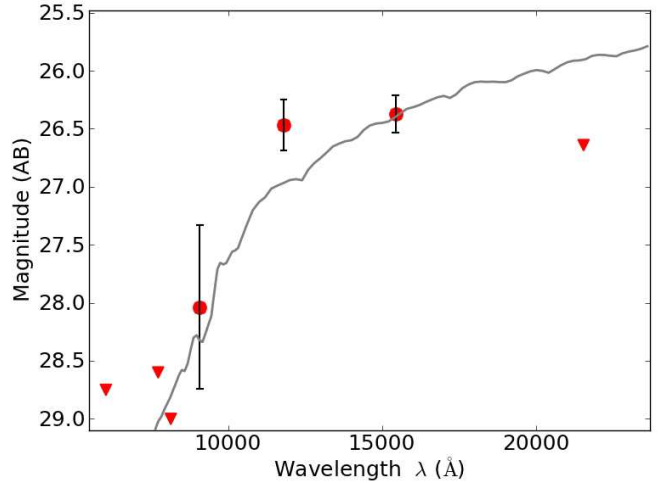}
\end{minipage}\end{minipage}
\end{center}
\caption{{\it (left)} Tests of a possible [\ion{O}{2}] line
  interpretation. An [\ion{O}{2}] line from another spectrum has been
  degraded to the S/N of our detected line and placed at slightly
  different wavelengths and in two different regions in an empty slit
  (around $9412\mbox{\AA}$ and in a place with fewer skylines for
  comparison). Due to the low S/N and the proximity of skylines we can
  not fully rule out the possibility of the line being an [\ion{O}{2}]
  based on this test {\it alone}. {\it (right)} Best fit SED when
  forcing the redshift to be $z<5$. Labels are as in Fig.~\ref{fig:phot}. The favored
  solution is at $z_{\rm{phot}}=1.4_{-0.6}^{+0.7} $ albeit with a very poor fit.}
\label{fig:o2test}
\end{figure*}

(1) Any of the three low redshift scenarios are strongly disfavored by
photometry. As shown in Figure~\ref{fig:o2test}-right, the fit
to the photometric data while forcing $z<5$ is very poor ($z_{\rm{phot}}=1.4_{-0.6}^{+0.7}$). Forcing it to $z=1.525$ (if line is
[\ion{O}{2}]), or accounting for the emission due to the line, does
not change the quality of the fit. The [\ion{O}{2}] doublet should
nominally be resolved at our resolution; however the sky emission at
these wavelengths degrades S/N and our ability to distinguish
individual components (Fig~\ref{fig:o2test}-left).
Unfortunately, in either scenario (Lyman-$\alpha$/[\ion{O}{2}]) no
other lines are expected in the wavelength range covered by the
spectrum ($7700\mbox{\AA}-10000\mbox{\AA}$). The line has the rest
frame EW of ${\sim}100\mbox{\AA}$ if at $z=1.5$ ([\ion{O}{2}] scenario)
and ${\sim}150\mbox{\AA}$ if at $z=0.43$ (H-$\alpha$). This is higher
than typical [\ion{O}{2}], and H-$\alpha$ emitters equivalent widths
as measured by \citet{straughn09}. Their median equivalent widths are
$36\mbox{\AA}$ and $73\mbox{\AA}$ for [\ion{O}{2}] and H-$\alpha$
respectively. Out of 30 [\ion{O}{2}] emitters only 3 have equivalent
width $>100\mbox{\AA}$. Hence, given the photometry and strength of
the line we conclude that the [\ion{O}{2}] scenario is very unlikely.

(2) [\ion{O}{3}] at $z= 0.880$
is even more strongly disfavored as we would expect to detect the
second line of [\ion{O}{3}] as well as H-$\beta$ line for typical line
ratios. That part of the spectra is clear, hence this identification
is ruled out by the data.  (3) H-$\alpha$ at $z=0.434$ is also
strongly disfavored by the photometry. 

Recently \citet{hayes12} cautioned against using photometry only to
select high redshift galaxies. They targeted a {$\J$}-band dropout from
\citet{laporte11} and discovered it was a low redshift
interloper. However, their situation is different from the one
reported here. While their target is a {$\J$}-band dropout by their
selection criteria, the object is detected blueward of Lyman-$\alpha$
in the {$\ii$}-band and the resulting best-fit model is poor. Even
without the {$\ii$}-band observations the photometry can be fit with a 
low-z template. In our
case the high redshift SED fit is good and the low redshift fit is
extremely poor.

We conclude that the line is most likely Lyman-$\alpha$ at $z=6.740
\pm 0.003$. The object has an AB magnitude of $26.5\pm0.3$ in the
{$\J$}-band (lensed, before correcting for magnification), which
corresponds to a continuum flux of
$f_{\lambda,\rm{c}}=3.3_{-0.8}^{+1.0}{\times}10^{-20}\mbox{erg/s/cm}^2/\mbox{\AA}$.
With an integrated line flux of
$f=(0.7\pm0.1\pm0.3){\times}10^{-17}\lunit$, the resulting line
rest-frame equivalent width is
$W^{\rm{rest}}(\mbox{Ly}\alpha)=f/f_{\lambda,\rm{c}}(1+z)=30^{+12}_{-21}\mbox{\AA}$
(Table~\ref{tab:drops}). The distribution of EW (including
non-detections) of the total sample of 10 dropouts to be presented in
a future paper will help distinguish between scenarios of Lyman-$\alpha$ opacity
around the epoch of reionization \citep{treu12}.

 Using the data we  also obtain rough estimate for the star
  formation rates (SFRs) using both Lyman-$\alpha$ and UV continuum
  luminosities, and Kennicutt's relations \citep{kennicutt98}.  We first estimate the
SFR from the Lyman-$\alpha$ luminosity $L_{\rm{Ly}\alpha}$ for the case B recombination theory as
$SFR_{\mbox{Ly}\alpha}=9.1{\times}L_{\mbox{Ly}\alpha}[M_{\sun}/\mbox{yr}]$, with $L_{\mbox{Ly}\alpha}$ in units of $10^{43}
\mbox{erg s}^{-1}$. We obtain $SFR{\sim}(3.3\pm1.5)/{\mu}\;M_{\sun}/\mbox{yr}$, where errors only include
measurement errors on $L_{\mbox{Ly}\alpha}$ and no uncertainties in
Kennicut's relation. Note that this is a
lower limit since $L_{\rm{Ly}\alpha}$ is not corrected for absorption effects
which depend on various parameters, including the neutral
fraction of the IGM and the kinematic status of neutral hydrogen. 

To convert UV luminosity $L_{\rm{UV}}$ into SFR we use
$SFR_{\rm{UV}}=1.4{\times}L_{\rm{UV}}[M_{\sun}/\mbox{yr}]$, where $L_{\rm{UV}}$ is in units of $10^{28}\mbox{erg s}^{-1}\mbox{Hz}^{-1}$. We estimate $L_{\rm{UV}}=(6.2_{-1.5}^{+1.8})10^{28}\mbox{erg s}^{-1}\mbox{Hz}^{-1}$ using $\J$-band
magnitude, as the central wavelength is very close
to rest-frame $1500\AA$, giving $SFR_{\rm{UV}}=8.7_{-2.1}^{+2.6}/{\mu}\:M_{\sun}/\mbox{yr}$. Both estimates are fully
compatible within systematic and statistical uncertainties implying  that the dust attenuation is very low
\citep{verhamme08}, which is also indicated by the blue  UV spectral slope
from $\J,\H$ and $\rm{K_s}$ band photometry. Dividing $SFR$ by magnification, the intrinsic star
formation rate is $SFR{\sim}2-3 M_{\sun}/\mbox{yr}$, consistent
with SFR from, e.g., \citet{labbe10}.

\section{Conclusion} \label{sec:conclusions} 

We have presented deep VLT spectroscopy of a strongly lensed
{$\z$}-band dropout galaxy behind the Bullet cluster. We detected an
emission line at $9412\mbox{\AA}$ with $>5\mbox{-}\sigma$
significance, which we identify as Lyman-$\alpha$ at $z=6.740\pm0.003$
at $>99$\%CL.  Correcting for magnification \citep[by a factor of
$\mu=3.0\pm0.2$ as discussed by][]{hall12,bradac09}, the intrinsic
(unlensed) line flux is $f=(0.23\pm0.03\pm0.10\pm0.02){\times}10^{-17}\lunit$ (Table~\ref{tab:drops}), which is ${\sim}2-3$
times fainter than the faintest spectroscopic detection of an LAE at
$z{\sim}7$
\citep{schenker12}.  Its intrinsic {$\H$}-band magnitude is $m^{\rm{int}}_{H_{\rm{160W}}}=27.57\pm0.17$, corresponding to an intrinsic
luminosity of $0.5L^*$ (where $L^*$ was calculated from the best fit
LBG luminosity function from
\citealp{bouwens11}).

The source is undetected in the four IRAC bands, which is not
surprising given that we would only be able to detect extremely red
galaxies with $m_{H_{\rm{160W}}}-m_{{\chone}}=4$ at 3-$\sigma$ in
$\chone$ for sources this faint. For comparison, $z=6.027$ source
behind A383 \citep{richard11} with an unusually mature stellar
population ${\sim}800 \mbox{Myr}$ and a multiply-imaged $z=6.2$ object
\citep{zitrin12} with a younger age $\sim180\mbox{Myr}$ have much
bluer colors $m_{H_{\rm{160W}}}-m_{\chone}{\sim}1.5$. Deeper Spitzer
data will be needed to probe the presence of mature stellar
populations in the galaxy we present here and other systems at high
redshift.

While this work presents only a single spectroscopic detection at
$z>6.5$, it nonetheless probes a very important region of parameter
space. As noted above, measuring the EW distribution of LAEs as a
function of redshift {\it and} luminosity is a very powerful tool to
study reionization, because the latter is likely the key factor
driving the trend of EW in luminosity \citep{dayal12}. The main
missing observational ingredient is a measurement of the EW
distribution for both luminous and sub-$L^*$ galaxies at the redshifts
of reionization.

Our source is the faintest one (in line flux) detected thus far and
is only the second firm spectroscopic detection of a sub-$L^{*}$
source at $z>6.5$. With future observations of dropouts magnified by cosmic
telescopes we plan to further increase this sample. Once completed,
this survey will help constrain the duration and physical processes
occurring at the epoch of reionization.

\acknowledgements We would like to thank the anonymous referee for
suggestions that greatly improved the paper and Sam Schmidt, Michele
Trenti, and Andy Bunker for stimulating discussions. Support for this
work was provided by NASA through HST-GO-10200, HST-GO-10863, and
HST-GO-11099 from STScI. EV acknowledges financial contribution from
ASI-INAF I/009/10/0 and PRIN MIUR 2009 ``Tracing the growth of
structures in the Universe: from the high-redshift cosmic web to
galaxy clusters''. TT acknowledges support from the NSF through
CAREER/NSF-0642621, through a Sloan Research Fellowship, and by the
Packard Fellowship. Part of the work was carried out by MB and TT
while attending the program "First Galaxies and Faint Dwarfs" at KITP
which is supported in part by the NSF under Grant No. NSF PHY11-25915.

\end{document}